\documentclass[12pt]{article}
\usepackage{epsfig}

\usepackage{amsmath}
\usepackage{amssymb}
\usepackage{euscript}

\begin{document}

% title
%%%%%%%%%%%%%%%%%%%%%%%%%%%%%%%%%%%%%%%%%%%%%%%%%%%%%%%%%%%%%%%%%%%%%%%%%%%%%%%
%%%%%%%%%%%%%%%%%%%%%%%%%%%%%%%%%%%%%%%%%%%%%%%%%%%%%%%%%%%%%%%%%%%%%%%%%%%%%%%
\begin{titlepage}

\begin{center}
{\LARGE\bf Radiative corrections to DIS$^{\star}$} \\
\end{center}

\vspace{15mm}

%\vspace*{6mm}
%\vspace*{6mm}
%\vspace*{6mm}
% authors

\begin{center}
{\large\bf  Mieczyslaw Witold Krasny } \\

\vspace{5mm}

LPNHE, Pierre and  Marie Curie University, Paris \\
and  \\
Institute of Nuclear Physics,  HNINP-PAS, Cracow

\end{center}

\vspace{20mm}
\begin{abstract}
%{\it Abstract:  
Early deep inelastic scattering (DIS) experiments  at SLAC discovered partons, identified them as quarks and gluons, and restricted the set of the candidate theories for strong interactions to those exhibiting the asymptotic freedom property. The next generation DIS experiments at FNAL and CERN confirmed the predictions of  QCD for  the
size of the scaling violation effects in  the nucleon structure functions. The QCD fits to their data resulted in  determining  the momentum distributions of the point-like constituents of nucleons. 
Interpretation of data coming from all these experiments and, in the case of the SLAC experiments, even
an elaboration of  the running strategies, would not have been possible without a precise understanding  of the electromagnetic radiative corrections. In this note I recollect the important milestones,
achieved in the period preceding the HERA era, in  the high precision calculations of the radiative corrections to DIS, and in the development of  the methods of their experimental control. I present  subsequently the 
measurement strategies and discuss  the  advanced radiative correction tools for the HERA experimental program,
with an emphasis on their role in the first, model independent, measurement of  the  partonic densities in the small-$x_{Bj}$ region.  These notes are dedicated  to Staszek Jadach  as a contribution to the celebration of his 60th-birthday.
\end{abstract}

\vspace{15mm}
\footnoterule
\noindent
{\footnotesize
$^{\star}$ Contribution to  the Cracow Epiphany Conference on LHC Physics, 4-6 January 2008, Cracow, Poland
}

\end{titlepage}

\section{Introduction}

DIS cross-sections, in particular those measured using the charged lepton beams, contain 
large contributions coming from the higher order QED processes. These contributions   
must be subtracted in order to interpret the measurements in terms of the nucleon structure functions. 

Let me recall the following two examples which illustrate the importance of the radiative corrections to DIS. 
The first one is the measurement of  the nucleon photo-absorption  cross sections for longitudinally and transversely polarized virtual photons. The kinematical dependence of their ratio, $R$, is driven by  the 
spin of the nucleon constituents. It is sensitive to the  diquark-quark structure
of the nucleons and, for the large four-momentum transfer $Q$,  to  the intrinsic transverse momenta of the quarks.
The  precision of measuring the kinematical dependence of $R$ was improved by the SLAC E140 experiment  \cite{E140}  to such an extent,    that  the overall measurement errors were no longer determined by the experimental ones,  but by the uncertainties in the size of the radiative corrections \cite{habilitation}. 
Another illustrative example  of the importance of radiative corrections to DIS   
is the comparison of the size of the QCD scaling violation effects and the radiative correction effects
in the proton structure function $F_2$ at small $x_{Bj}$. In this region the effects of the radiative corrections are particularly large - 
they are of the same magnitude as those reflecting a change by 200 MeV of the  $\Lambda_{QCD}$  value  \cite{krasnyplenary}. 

What is  the main reason for such a large sensitivity of the DIS observables  to the size of the radiative 
corrections? In each of  the early electron-beam DIS scattering experiments at SLAC and DESY, as well as in the next generation muon-beam DIS  experiments at CERN and FNAL,  the kinematic variables were reconstructed 
using the initial lepton energy, the scattered lepton momentum and its scattering angle.
Within  such a reconstruction scheme the {\it apparent} value of the four-momentum 
transfer from the lepton to the nucleon, $Q_r $, could differ significantly from the {\it true} four-momentum 
transfer,  $Q_t $,  in the presence of untagged, hard photon radiation  processes. Due to a  steep rise 
of the DIS cross section in the region of small $Q_t$ the contribution of the hard-photon radiative processes were 
found to be of similar size as the contribution of the Born process.  This "kinematical bias"  was the
dominant  source of the large radiative corrections. 

At HERA, where the leptonic radiative corrections were expected to be even larger,  
three complementary strategies of reducing their impact on the overall measurement precision were planned 
\cite{krasnyplenary} and realized \cite{RadCor95}. The  first one 
was to apply the same correction procedure   as for  the  early SLAC experiments. This  
procedure required a  precise modeling  of the Born  cross section in the small $Q_t^2$ region,
and the  development of   dedicated methods of its experimental control, 
e.g.  by using the Compton events \cite{Courau}.
The second one was to use the capacity of the HERA detectors in direct (zero-angle photon 
calorimeter),  or indirect (hadronic energy flow) detection of 
hard photons radiated in the angular region collinear to  the incoming electron direction.  The third one
was to determine the event kinematic variables using (entirely  or partially) the hadronic final state observables.

Each of  the above three methods required  high-precision calculations of the radiative corrections and their subsequent implementation in a  form which was easily applicable to the data analysis procedures. 
This note brings into light important milestones in this domain. 
It tries to pay a tribute to those  who provided  indispensable tools for the DIS experimental program. 
Their painstaking  work, often hardly  visible in the bulk of the published experimental results 
was of primordial  importance for the analysis of the DIS data. It enabled a precision mapping  
of  the partonic distribution functions -  a base for understanding 
the wide-band-partonic beams which will be used by  the LHC experimental program.

\section{The tools and the methods for the SLAC, FNAL and CERN  DIS experiments}

The early SLAC experiments  were confronted with the following three challenges.
Firstly, in order to increase the observed event rate,  the collision data were collected with relatively thick 
(couple-of-percent-radiation-length-long) targets. Therefore, the radiative corrections to DIS (internal radiative corrections) had to be calculated simultaneously  with the target-length-dependent external radiative corrections.
In  the SLAC kinematic region  the effect of internal radiative corrections was roughly 
of the same magnitude as the effect of the external ones. The calculations of the latter 
was controlled experimentally by using  the variable thickness targets \cite{KendalFriedman}.
Secondly, the deep inelastic cross section for absorption of transverse and longitudinal 
photons were not known at the time when the first measurements were done. The unfolding of the 
radiatively  corrected observables  required thus a global  iterative procedure, and could not be confined to the local correction factors. Thirdly, and perhaps most importantly, the available computing power was limited and several approximations had to made while calculating the size of radiative corrections.

Given all these constraints,  the early DIS experiments 
at SLAC were using various peaking approximations 
based on the formulae of L.W. Mo and Y.S. Tsai \cite{MOTSAI}. The combined effect of the external 
and the internal radiative corrections was calculated using the equivalent radiator method in which the 
radiative corrections were approximated by adding  two hypothetical radiators, 
of the $Q^2$-dependent length, one in front  and one
behind  the vertex position. In the numerical 
integration procedures the energy peaking approximation was used. Such an 
approximation was indispensable to reduce  the integration domains
to the strips along the energies of the initial and the scattered electron. 

For the next generation SLAC experiments, and 
for the muon-beams CERN and FNAL experiments, the above approximations 
turned out to be the dominant overall precision-limiting factors. Therefore,  new methods had to be developed. 
From the theoretical side, the most important development for the  
precision DIS program  came from D.  Bardin and his Dubna group \cite{Bardin}.  They  calculated the full set of internal radiative corrections including the hadronic corrections, the electroweak effects,  and the  soft photon exponentiation effects. These calculations, implemented within their  TERAD86  program,  provided a very important tool to verify the precision of several  approximations, present in the calculations of  the internal radiative corrections for the SLAC  DIS experiments.   
Several precision-limiting approximations  in the 
procedures based on the Mo-Tsai formulae were identified \cite{krasnyslac, sridhara}. In addition,  the increased power of the SLAC computing facilities allowed avoidance of the approximations 
made in the calculations of  the combined effects of the  internal and external corrections. The improvement in  the  precision of the 
radiative corrections achieved using new calculation procedures \cite{krasnyslac} triggered the  reanalysis of the early SLAC DIS data  resulting in  a consistent picture of the structure functions measured at SLAC and at CERN.

\section{The tools and the methods for the HERA DIS program}

For the HERA research program,  novel theoretical tools and novel experimental methods of 
controlling the processes of hard photon emission had to be developed in order to cope with large  radiative 
corrections.

Already before the startup of HERA,  
experimentalists were well equipped with the new 
generation  computer-programs for radiative corrections (both for the analytical calculations 
and for the Monte-Carlo generation of radiative processes)  \cite{spiesbergerplenary}, and with the HERA-specific methods of application of these corrections to the experimental data \cite{krasnyplenary}.

The TERAD91 program by D. Bardin et al. \cite{Bardin91} had its roots in the TERAD86 program applied 
successfully to  the earlier DIS experiments. It included, as well,  the code for the analytical calculations
of the complete electroweak corrections to the NC and CC scattering in the quark-parton model
(DISEPNC and DISEPCC). The EPRC91 program by H. Spiesberger et al. \cite{Bohm},  was a package of programs for the calculations of complete electromagnetic and weak corrections to the DIS NC and CC scattering. The above two programs were tested using  identical input structure functions and shown to agree with each other at  the level below  1\%. 

The HERA experiments H1 and ZEUS  were the first DIS experiments capable of measuring  
the  hadronic flow associated with the CC and NC collisions. This allowed us to introduce novel reconstruction methods for the DIS  kinematical variables and novel methods of applying the radiative corrections to the measured 
observables. One of the  most important aspect of  these methods was to go beyond the classical scheme, in which the radiative corrections were applied at the last analysis step -   i.e. 
after correcting for the measurements effects -  by introducing the  methods in which unfolding  of the experimental effects was done simultaneously with correcting the data for radiative effects. In order to implement  such  methods the development of the Monte Carlo generators involving radiative processes was indispensable.

The first attempt in this direction was the LESKO-C generator  by S. Jadach \cite{LESKOC}.
This  Monte Carlo generator was based on the collinear approximation for the processes of bremsstrahlung  from the polarized initial lepton. This program evolved to the  LESKO-F program \cite{LESKOF}.  It included 
the complete $O(\alpha)$  QED radiative corrections to the lepton line for the NC deep-inelastic scattering. 
Later, in the LESKO-YFS generator, by W. Placzek and S. Jadach,   the multi-photon radiation processes were added \cite{YFS}. 
An interface of these programs to the LEPTO \cite{LEPTO} program for parton cascades and fragmentation, the FRANEQ program by W. Placzek\cite{FRANEQ}, allowed the generation of  complete final states.

The main limitation of these  generators  was that they were applicable 
only to the processes of large four-momentum transfer to the hadronic system, $Q_t$ - for which protons could  be considered as composed of quarks and gluons. Since the reconstruction of this quantity required  modeling of perturbative and non-perturbative QCD effects,  and was sensitive to a precise modeling of the detector response
to the produced hadrons,  this program had a limiting power for high-precision unfolding of the radiative corrections
to electron inclusive observables.  Nevertheless,  these generators were of great importance in the early phase of development of the experimental methods of controlling the radiative processes at HERA \cite{Jezabek, Krasny-Placzek} and while developing a novel method of measuring the proton longitudinal structure function  \cite{Fl}.

The most important and the most widely used Monte-Carlo generator for the HERA experimental 
program, HERACLES,  was developed by H. Spiesberger et al. \cite{SpiesbergerHeracles}. The HERACLES 
program allowed a separate treatment of the Born term and the  radiative terms (soft and hard 
bremsstrahlung and the corresponding virtual corrections). It included the  Compton part and quarkonic
radiation. Users were allowed to use any parameterization of the input structure functions including 
the longitudinal structure function. For those of users for whom the quark-parton modeling of the proton structure was sufficient, the interface DJANGO to  the LEPTO and JETSET programs \cite{LEPTO}
was provided.
 
The HERA experiments were the first DIS experiments in which the  radiative hard 
photons could be identified and measured \cite{RadCor95}.
Events in which photons were emitted in the angular range of $\sim$ 0.45 mrad with respect to the 
incoming electron direction,  and events in which the radiative photon 
transverse momentum balanced (to a similar precision) that of the scattered electron,  were used
to measure the machine luminosity, and to monitor the beam position and divergence  at the HERA interaction points. 
Direct and indirect (using the hadronic flow observables) detection of 
radiative photons allowed an experimental cross-check of the size of radiative corrections and 
to measure the proton structure functions in the kinematical regions not accessible to the standard methods.

\section{Radiative corrections for  the first HERA data}

One of the first and most cited  HERA result  based on the data collected in the 1992 run 
was the first measurement of the proton charge-structure in the small-$x_{Bj}$ region \cite{KrasnyF2}. 
The observed strong rise of the differential cross section, if  interpreted in terms of 
the leading twist partonic distributions indicated that, in the small-$x_{Bj}$ 
region,  the  proton momentum was  carried mostly by  gluons. 
The H1 collaboration strategy for the  first measurement at HERA of the proton structure  \cite{KrasnyF2} anticipated 
that both the physics,  and the detector and machine performance,   could not be precisely modeled at the time of the 
initial measurement. 

Therefore,  dedicated measurement methods based on a direct control of the experimental errors and on the factorization of the radiative correction effects from the detector performance effects had to be invented. The important aspect of the 
initial measurement strategy  was to determine the differential cross sections using independent  methods having different sensitivity not only to the size of radiative corrections but also to the assumed shape of the differential cross-section in the unmeasured region. Such a strategy, similar to the dedicated strategies of the first DIS experiments at SLAC, assured a precise control of the radiative corrections 
in spite of the fact that,  for the measurements based on the electron variables, they were exceeding 100\% of the measured
values.  The implementation of such strategies was possible owing to the effort of the 
DESY radiative correction group\footnote{The DESY radiative correction group  was convened by Hubert Spiesberger
and by the author of this note.} involving  both the theorists and the experimentalists who  prepared 
highly efficient  tools and strategies.

\section{Conclusions}

Precision measurements in high-energy physics, as long as they require modeling of the underlying phenomena 
using  the framework of  the perturbative quantum field theories,  will always need a symbiotic  effort of theorists and experimentalists.
One of the most obvious targets for such a collaborative work  is to prepare the high-precision  theoretical tools to control the higher-order radiative corrections. Their  "user-market" quality is  measured presently by a simple criterion: to which extent  such tools encapsulate their sophisticated  technical aspects  while being robust and versatile.  
The tools for the DIS program fulfilled perfectly this  quality requirement. The measure of their success is, paradoxically, that they became, being  uncontroversial, invisible in the results of the HERA experiments. 
The challenge for elaboration  of advanced  EW and QCD radiative correction tools for the LHC experimental program is, first of all, to preserve the interest of theorists to work in such a "shadow" activity which is as much unexposed as it is useful for the community of experimentalists. Staszek Jadach's group is one of the  groups of ``last Mohicans"  staying in this field. What must be stressed is that, for a high -precision scrutiny of the Standard Model at LHC,  these tools must 
be elaborated in an undissociated way with preparation of the dedicated measurement strategies in which 
the theoretical (modeling) uncertainties  and the experimental errors could be controlled independently. 
The latter aspect, of lesser importance for the LEP and HERA programs,  is of extreme importance and urgency at LHC 
where the modeling of the partonic wide-band-beams and higher-order 
QCD correction involves a sufficient freedom to "swallow" the experimental 
manifestation  of a wide class of  novel phenomena.

\end{document}